\documentstyle[preprint,pre,aps,eqsecnum]{revtex}
\begin{document}
\preprint{PISA IFUP-TH XX, June 1993}
\draft
\title{
The Kramers equation simulation algorithm:\\
II. An application to the Gross-Neveu model
}
\author{Matteo Beccaria$^{(1,2)}$, Giuseppe Curci$^{(2,1)}$,
Luca Galli$^{(1)}$}
\address{
(1) Dipartimento di Fisica, Universit\'a di Pisa\\
Piazza Torricelli 2, I-56100 Pisa, Italy\\
(2) I.N.F.N., sez. di Pisa\\
Via Livornese 582/a, I-56010 S. Piero a Grado (Pisa) Italy
}
\date{\today}
\maketitle
\begin{abstract}
We continue the investigation on the applications of the Kramers
equation to the
numerical simulation of field theoretic models.
In a previous paper we have described the theory and proposed various
algorithms. Here, we compare the simplest of them with the
Hybrid Monte Carlo algorithm studying the two-dimensional lattice
Gross-Neveu model. We used a Symanzik improved action with
dynamical Wilson fermions. Both the algorithms allow
for the determination of the critical mass. Their performances
in the definite phase simulations are comparable with the Hybrid Monte Carlo.
For the two methods, the numerical values of the measured quantities
agree within the errors and are compatible with the theoretical
predictions; moreover, the Kramers algorithm is safer from the point of
view of the numerical precision.
\end{abstract}
\vskip 1truecm
\pacs{PACS numbers: 11.15.Ha, 05.50.+q, 11.30.Na, 11.30.Rd}
\section{Introduction}
In a previous work~\cite{previous}, we used a compact operatorial
formalism to study
the applicability of the Kramers equation to the numerical simulation
of quantum field theories
on the lattice.
In the standard Monte Carlo approach, we must generate ensembles of
states distributed
according to a given statistical weight.
In~\cite{previous}, our strategy has been to realize a diffusive process
involving the physical degrees of freedom but also auxiliary
variables which drive the diffusion. The final equilibrium distribution
approximates
the desired one. The resulting schemes can be made exact by a global
accept-reject test
and provide thus generalizations
of the usual Hybrid Monte Carlo~\cite{hmc1,hmc2} and of the exact hyperbolic
algorithm~\cite{horowitz2}.
For these algorithms, a good behaviour with increasing volume
and from the point of view of numerical precision is expected.
Moreover, in~\cite{previous}
we have shown how the freedom in the auxiliary variables sector
may be used to reduce
significantly the autocorrelations in the free field case.
In this paper we study the behaviour of the simplest among the
proposed schemes. We take
the Hybrid Monte Carlo as a reference point.
Our theoretical laboratory is the two-dimensional $N$-fermion
Gross-Neveu model with
Wilson fermions~\cite{curci}.
This model is an asymptotically free purely fermionic model with a
discrete $\gamma_5$
symmetry whose dynamical breakdown leads to mass generation. It has an
exact large $N$
solution and is $1/N$ expandable~\cite{expansion} allowing for analytical
investigation of its
non perturbative properties.
In our lattice simulation of the model we use the Wilson formulation to
avoid the doubling problem and let the fermion number $N$ be free.
However, using Wilson fermions,
a bare mass must be introduced and tuned in order to restore the
chiral symmetry in the continuum limit. The restoration of chiral
symmetry is a
significant test for new algorithms: previous works~\cite{curci,rossi}
showed that non exact algorithms (Langevin, pseudofermions) fail to
locate the critical
point by means of mixed-phase techniques. In this work we shall show
that the algorithm
based on the Kramers equation does not encounter this problem.
The use of two-dimensional four-fermions models as prototypes
for realistic theories like $QCD$ is somewhat limited.
However, a non trivial model with a discrete chiral symmetry
breaking and many dynamical fermionic degrees of freedom implies enough
problems to be
a severe test for our algorithms.

In Sec.~\ref{GNmodel} we introduce the Gross-Neveu model.
In Sec.~\ref{algorithms} we describe explicitly the algorithms that we
have used.
In Sec.~\ref{comparison} we discuss how we compared them.
In Sec.~\ref{runs} we review the previous simulation of the Kramers
algorithm and of the
Symanzik improved Gross-Neveu model. We then describe our explicit numerical
simulation and present the results.
Finally, Sec.~\ref{summary} is devoted to a short summary and conclusions.
\section{The Gross-Neveu model}
\label{GNmodel}
The action of the model in the continuum is
\begin{equation}
S = \int d^2x \left({\bar\psi}^{(\alpha)}
{\partial \!\!\! /}
\psi^{(\alpha)}-\frac{1}{2} g^2
\left({\bar\psi}^{(\alpha)}\psi^{(\alpha)}\right)^2 +
m{\bar\psi}^{(\alpha)}\psi^{(\alpha)}\right) \qquad \alpha = 1\cdots N.
\end{equation}
where $\psi^{(\alpha)}$ is a multiplet of two-dimensional Dirac fermions.

Rewriting the quartic interaction with a Lagrange multiplier we obtain
\begin{equation}
S = \int d^2x \left({\bar\psi}^{(\alpha)}
{\partial \!\!\! /}
\psi^{(\alpha)}+(\sigma+m)
{\bar\psi}^{(\alpha)}\psi^{(\alpha)}+\frac{1}{2g^2}\sigma^2
\right).
\end{equation}
The manifest $U(N)$ symmetry of the model can be enlarged to $O(2N)$ by
writing the fermionic fields
in terms of their hermitean Majorana components. We refer to
\cite{counterterm}
for a discussion of the renormalization
properties of the continuum model in the two
formulations with and without the auxiliary $\sigma$ field.
In the chiral limit the model enjoys the discrete symmetry
\begin{equation}
\psi \to \gamma_5\psi \qquad
{\bar\psi} \to -{\bar\psi}\gamma_5 \qquad
\sigma \to -\sigma.
\end{equation}
As a consequence, the potential of $\sigma$ , related to that of the
composite
field~${\bar\psi}\psi$, is symmetric under the exchange
$\sigma\to -\sigma$ and possesses two degenerate minima.
The chiral symmetry is spontaneously broken. The same holds on the
lattice, but
in the Wilson formulation we need an explicit mass term
to control mass renormalization.

The $1/N$ expansion of the model
is obtained by integrating out the fermion fields.
By introducing the large-$N$ coupling $\lambda = N g^2$ and the
field $\Sigma = \sigma + m$, the corresponding
``effective'' action is
\begin{equation}
\label{effaction}
S_{eff} = N \int d^2x \left(\frac{1}{2\lambda} (\Sigma-m)^2 - \mbox{Tr}\log
M(\Sigma)\right) \ \ \ \ \ M(\Sigma) =
{\partial \!\!\! /}
 + \Sigma .
\end{equation}
In a perturbative expansion in powers of $g$, the fermions
remain massless in the chiral model.
However, already in the leading order of the $1/N$ expansion, a non-zero
expectation value of the $\Sigma$ field is dynamically generated
playing the role of a non-perturbative fermion mass. From the study of
the phase diagram of the
model we can infer the existence of a critical point in the weak
coupling region. At this
point the chiral symmetry can be recovered (and dynamically broken)
with an appropriate choice
of the bare critical mass. The vacuum expectation value
$\langle \Sigma \rangle$ is exponentially depressed as $\lambda\to 0$
following
the leading renormalization group prediction $\langle \Sigma \rangle
= \pm \Lambda \exp(-\pi/\lambda)$. Actually, on a finite lattice at non
zero $\lambda$
one chooses the perturbative mass in order to have a double well shaped
potential
for the $\Sigma$ field having two degenerate minima. The $\Sigma\to-\Sigma$
symmetry is
recovered only in the chiral limit; this is reminiscent of the doubling
problem since one has
exactly
\begin{equation}
\langle F(\psi, \bar{\psi}, \Sigma, r)\rangle =
\langle F(\gamma_5\psi, -\bar{\psi}\gamma_5, -\Sigma, -r)\rangle,
\end{equation}
where $r$ is the Wilson (or Symanzik) coefficient of the mass term
which solves the doubling.
Quantities like the asimmetry in the steepness of the
potential in the two minima can be $1/N$ expanded and computed.

Since we are in two dimensions, finite volume effects are particularly
dangerous. Therefore
we improved the lattice Wilson action at the tree-level following the
method of
Symanzik~\cite{symanzik}. The resulting action improved at order $O(a^2)$ is
\begin{eqnarray}
\label{model}
S &=& \sum_{n\mu} {\bar\psi}_n\left[\gamma_\mu(\frac{2}{3}(\psi_{n+\hat{\mu}}
-\psi_{n-\hat{\mu}})-
\frac{1}{12}(\psi_{n+2\hat{\mu}}-\psi_{n-2\hat{\mu}})) + \right. \nonumber\\
&& \left. + \frac{r_s}{2}(\psi_{n+2\hat{\mu}}-4\psi_{n+\hat{\mu}}+6\psi_n
-4\psi_{n-\hat{\mu}}+\psi_{n-2\hat{\mu}})\right] + \\
&&
+ \sum_n \left[{\bar\psi}_n\psi_n (m+\sigma_n) + \frac{N}{2\lambda}
\sigma_n^2\right], \nonumber
\end{eqnarray}
where we have understood the flavour indices. The Fermi fields have
antisymmetric boundary conditions
in the temporal direction.
The parameter $r_s$ is the Wilson parameter for the
Symanzik action. We have taken $r_s = 1/3$ remarking that
the strong coupling matching between Wilson and Symanzik actions is
at $r_s = 1/3\  r_w$.
In~\cite{curci} the interested reader can find a detailed account of the
leading and next-to-leading $1/N$ expansion of the model described
by Eq.~(\ref{model}).
In momentum representation, the fermionic matrix is
\begin{equation}
\label{explmatrix}
M = i\gamma_\mu\bar{p}_\mu + M_S(p) + \Sigma,
\end{equation}
where we have defined
\begin{eqnarray}
M_S(p) &=& \frac{1}{2} r_s \sum_\mu\hat{p}_\mu^4, \nonumber \\
\hat{p}_\mu &=& 2\sin\frac{p_\mu}{2}, \\
\bar{p}_\mu &=& \sin p_\mu \left( 1 + \frac{1}{6}
\sum_\mu\hat{p}_\mu^2\right), \nonumber
\end{eqnarray}
On a finite $L\times T$ lattice, the form Eq.~(\ref{effaction})
of the effective action
together with Eq.~(\ref{explmatrix}) allows for the computation
of any relevant quantity in
the leading $1/N$ limit.
\section{Description of the algorithms}
\label{algorithms}
We now describe in an unified scheme the algorithms that we have used.
We give all the details showing how the fermionic fields are dealed with.
Because of the particular structure of $M(\Sigma)$, we may introduce
$N$ real auxiliary fields $\chi^{(\alpha)}$ ($\alpha = 1\dots N$ and Dirac
indices understood)
and conjugate bosonic momenta $\pi$ to form the extended action
\begin{equation}
S = \sum_{sites}\left(N \ \frac{(\Sigma-m)^2}{2\lambda} + \frac{1}{2}\pi^2+
\frac{1}{2} {\chi}^{(\alpha)}
\left[M(\Sigma)^T M(\Sigma)\right]^{-1}
\chi^{(\alpha)}
\right).
\end{equation}
For each given configuration of $\Sigma$ and of the auxiliary
pseudofermionic fields $\chi$,
we define the following $N$ fields
\begin{equation}
\Phi^{(\alpha)}(\Sigma) = \left[M(\Sigma)^T M(\Sigma)\right]^{-1}
\chi^{(\alpha)},
\end{equation}
where the sparse matrix inversion is performed by means of the
Conjugate Gradient algorithm

The $\chi$ fields action is quadratic. Therefore,
we can extract randomly the $\chi$ fields according to their
exact equilibrium distribution and evolve the other fields between
two successive such refreshments.
Note that this procedure
introduces another free parameter, namely the frequency of the $\chi$ fields
update.

The single sweep updating is given by the following algorithm~\cite{previous}
\begin{itemize}
\item
Refresh the auxiliary $\chi^{(\alpha)}$ according to
\begin{equation}
\left\{
 \begin{array}{l}
  \chi^{(\alpha)} = M^T(\Sigma)\ \eta^{(\alpha)}, \\
  \pi = \eta .
 \end{array}
\right.
\end{equation}
Here, $\eta^{(\alpha)}$ and $\eta$ are vectors of gaussian random
numbers with zero
mean and unit variance. This step must be done only every $k$ sweeps.
\item
Update the momenta $\pi$ according to
\begin{equation}
\pi^\prime = e^{-\gamma \epsilon} \pi + \sqrt{1-e^{-2\gamma\epsilon}} \ \xi ,
\end{equation}
where $\xi$ is a gaussian random number with zero
mean and unit variance. The parameters $\gamma$, $\epsilon$
are positive real numbers.
\item
Integrate the equations of motion associated with the force
\begin{equation}
F_n(\Sigma)=
-\frac{N}{\lambda}(\Sigma_n-m) + \frac{1}{2}\ \Phi^{(\alpha)}
\frac{\delta}{\delta\Sigma_n}\left(M^T(\Sigma)M(\Sigma)\right)\Phi^{(\alpha)}
\end{equation}
($n$ is a site index) using $N_{md}$ iterations of the leap-frog scheme
\begin{eqnarray}
\pi(t+\epsilon/2) &=& \pi(t)+ \varepsilon/2\ F(\Sigma(t)), \\
\Sigma(t+\epsilon) &=& \Sigma(t) + \varepsilon \pi(t+\epsilon/2), \\
\pi(t+\epsilon) &=& \pi(t+\epsilon/2) + \varepsilon/2\ F(\Sigma(t+\epsilon)),
\end{eqnarray}
or its dual form in which the roles of $\Sigma$ and $\pi$ are interchanged.
\item
Perform a Metropolis test between the states before and after the
integration of
the equations of motion. If the test fails, reject the proposal
 for the $\Sigma$ field and
negate all the momenta
\begin{equation}
\Sigma\to\Sigma_{old}\qquad\qquad\pi\to -\pi_{old}
\end{equation}
otherwise accept the proposal for both $\Sigma$ and $\pi$.
\end{itemize}
We have four free parameters: $k$, $N_{md}$, $\epsilon$ and $\gamma$.
Detailed Balance is exactly satisfied for each set of values they take.
The algorithm under study (which we shall call ``Kramers algorithm'') is
obtained with $N_{md} = 1$ and arbitrary values of $k$, $\gamma$
and $\epsilon$.
The usual Hybrid Monte Carlo corresponds to $k=1$, $\gamma\to\infty$ and
arbitrary values of $N_{md}$, $\epsilon$. Of course, in
the Hybrid Monte Carlo case, there is no mixing between the old and
new momenta
and it is unnecessary to negate them on
rejection.

We remark that from a physical point of view, the Kramers algorithm
possesses two distinct time
steps, $\epsilon_{irr} = \epsilon\gamma$
drives the ``irreversible'' motion, whereas $\epsilon_{rev} = \epsilon$
drives
the ``reversible'' one.
Apart from effects coming from the negation of momenta, we expect that
the Kramers and the
Hybrid Monte Carlo algorithms to behave similarly when
$\epsilon_{irr}\sim 1/N_{md}$ with $\epsilon_{rev} = \epsilon_{md}$.
\section{Comparison}
\label{comparison}
In~\cite{previous} we argued that the Kramers algorithm should behave
better than the Hybrid
Monte Carlo for large volumes. The argument is rather naive and must
be confirmed numerically.
Consider a lattice model satisfying $L \gg \xi$, where $L$ is the
lattice size and $\xi$
the correlation length. On general grounds, the acceptance rate of the
Hybrid Monte Carlo is
known to be~\cite{hmc2}
\begin{equation}
P_{acc} \sim \mbox{erfc}(c N_{md} \epsilon^3 \sqrt{V}).
\end{equation}
Optimal tuning is expected to require $N_{md} \sim 1/\epsilon$ giving a
sweep-sweep
correlation $\tau\sim 1/\epsilon$.
If the volume is varied, we have to scale $\epsilon \sim V^{-1/4}$ in
order to keep the
acceptance probability constant. The advantage of the Kramers algorithm
is that $N_{md} = 1$. If
we neglect the influence of the momenta negation on the acceptance, then
the scaling relation is modified to $\epsilon \sim V^{-1/6}$.
If, moreover, the optimal
autocorrelations stay $\sim 1/\epsilon$ then it follows that on large
volumes the Hybrid Monte
Carlo behaves worst than Kramers.

Empirically the maximum acceptable value for $\epsilon$ in the
Kramers algorithm was rather greater than the one of the Hybrid Monte Carlo.
This is reasonable
because, at fixed $\epsilon$,  as $N_{md}$ is increased the acceptance rate
decreases before reaching a plateaux.

Another advantage of the Kramers algorithm is related to the numerical
precision,
the opportunity of having $N_{md} = 1$ without penalty is surely welcome;
for instance,
in $QCD$, as
the volume ($L^4$) grows or the quark mass ($m$)
is decreased we must have $N_{md} \sim L/m^{3/4}$ and some protection
seems necessary
to protect irreversibility against the accumulation of numerical errors.

To compare quantitatively the performances of the
two algorithms we must determine which is the computer time needed to
have a given
statistical error. If we denote with $\tau_\Omega$ the integrated
autocorrelation of
the observable $\Omega$, then the computational
cost of an algorithm measuring $\Omega$ is proportional to
$2\tau_{\Omega}$ which is
the reduction factor which determines the actual number of independent
measures.
This factor must be multiplied by the number of Conjugate Gradient
inversions needed to
make a single sweep. Indeed, in a model with dynamical fermions it is
reasonable to neglect the time expended during all the other steps of
the algorithm.
In our scheme we have blocks each made of $N_{md}$ molecular dynamics
steps followed by a
Metropolis test and we refresh the pseudofermionic fields $\chi$ only
at the end
of $k$ such blocks. Let us call $LF_1$ and $LF_2$ the two dual second
order leap-frog schemes
(see~\cite{previous} for a compact review of symplectic integrators)
which we sketch as follows
\begin{eqnarray}
LF_1 &:& \pi(t_0) \to \Sigma(t+\epsilon/2) \to \pi(t+\epsilon), \\
LF_2 &:& \Sigma(t_0) \to \pi(t+\epsilon/2) \to \Sigma(t+\epsilon). \nonumber
\end{eqnarray}
Keeping $\chi$ fixed,  a new Conjugate Gradient
inversion is needed only when the $\Sigma$ field configuration or
the $\chi$ fields change.
The advantage of the $LF_1$ scheme is that it computes the old and
new action at the
same evolution times at which the force itself must be evaluated.
The quantity determining the computational cost is
the number of inversions per sweep computed between two refreshment of $\chi$
\begin{eqnarray}
\label{funny}
{\cal C}(k, N_{md})_{LF_1} &=& N_{md} + 1/k, \\
{\cal C}(k, N_{md})_{LF_2} &=& N_{md} + 1 + 1/k, \nonumber
\end{eqnarray}
and the actual performance is just this ($\Omega$ independent) factor
corrected by the
($\Omega$ dependent) autocorrelation. From~Eq.(\ref{funny}) we see
that $LF_2$ must certainly
be ruled out at small $N_{md}$. For what concerns $k$,
in the Hybrid Monte Carlo case the $1/k$ term is usually not very
important because $N_{md}\gg 1$.
However, with the Kramers algorithm,
a large $k$ is twice better than $k=1$. Moreover $k>1$ may allow
for a more efficient
exploration of the phase space.

We remark that our conclusions hold provided that $LF_1$ and $LF_2$ have
the same behaviour
concerning autocorrelations.
\section{Simulation and numerical results}
\label{runs}
\subsection{Review of previous results}
\subsubsection{Previous study of the Kramers algorithm}
In~\cite{horowitz2}, the Kramers algorithm was applied to the simulation of
quenched compact $QED$ on a $8^4$ lattice in the disordered phase.
Measures of the
average plaquette and of the average $2\times 2$ Wilson loop were
made using also the
Hybrid Monte Carlo algorithm as a reference method. The performances
of the two algorithms
turned out to be comparable.
\subsubsection{Previous study of the improved Gross-Neveu model}
In~\cite{curci}, the Symanzik improved Gross-Neveu model was studied on
the lattice
using the pseudofermion
algorithm. In this scheme the variation of the fermionic contribution
to the action is
expanded to the lowest order in the $\Sigma$ field variation $\delta\Sigma$.
\begin{equation}
\delta S \to \frac{N}{2\lambda}\ (\Sigma-m)^2\ \delta\Sigma +
\frac{N}{2\lambda}\
\delta\Sigma^2 - N\ \mbox{Tr} M^{-1}(\Sigma)\ \delta\Sigma .
\end{equation}
The fermionic matrix inversion was performed computing
\begin{equation}
M_{ij}^{-1} = \langle \chi_i\chi_j^\dagger\rangle
\end{equation}
as a Monte Carlo average in presence of the quadratic action
$S_\chi=\chi^\dagger M^{-1}\chi$.
The systematic error to be kept under control has two contributions.
It gets an easily
controllable statistical contribution from the inversion and a
difficult contribution
due to the truncation of the fermionic term at small but non
zero $\delta\Sigma$.
An extrapolation as $\langle (\delta\Sigma)^2\rangle\to 0$
must be performed. Using the mixed-phase tecnique, the critical bare
mass $m_c$ was
determined at different values of the parameters and of the step
$\langle (\delta\Sigma)^2\rangle$. The extrapolation
did not agree with the theoretical prediction in the case of
$\langle \Sigma\rangle$ and was explained
as a wrong determination of $m_c$. Moreover, deviations from the
Schwinger-Dyson equations of the
model did not show the expected linear dependence on
$\langle (\delta\Sigma)^2\rangle$.
These problems were absent for $\langle \Sigma^2\rangle$ which is expected to
have a milder dependence on $m_c$.

In~\cite{rossi}, the same kind of measurements were done using two
different algorithms: the Langevin
one with bilinear noise and the Hybrid Monte Carlo. In both cases, the
inversion of the fermionic matrix
was performed by the Conjugate Gradient algorithm.
The Langevin algorithm is not exact, given a step $\epsilon$ the
update of the bosonic field
is computed as
\begin{eqnarray}
\Sigma(x,t+\epsilon) &=& \Sigma(x,t) -
\epsilon \frac{\delta S}{\delta\Sigma(x,t)} +
\sqrt{\epsilon}\ \eta(x,t) + \\
&& + \epsilon \sum_{y,z,w}\ \eta_f(y,t)\ M^{-1}_{yz}\
\frac{\delta M_{zw}}{\delta\Sigma(x,t)}
\ \eta_f(w,t), \nonumber
\end{eqnarray}
where $\eta$ and $\eta_f$ are gaussian fields normalized as
\begin{equation}
\langle \eta(x, t)\eta(x^\prime, t^\prime)\rangle =
2\delta_{xx^\prime}\delta_{tt^\prime} .
\end{equation}
An extrapolation as $\epsilon\to 0$ is needed.
The Langevin algorithm was found to be completely unable to determine
the critical mass since the
critical point was so unstable that different seeds of the random number
generator gave
different results. Using the theoretical value for the critical mass
the simulation in the definite
phase was stable only for the smallest values of $\epsilon$. On the
other hand, the Hybrid Monte
Carlo simulations turned out to be perfectly stable in both the mixed
and definite phase simulations.
The results (included the stable Langevin runs) were consistent with
the theoretical predictions.
\subsection{Our results}
In this work, we have repeated the measures of~\cite{curci,rossi} using our
favoured algorithm. Moreover we have studied the variation of the
performance as the
tuning is varied, an information which is very important from a
practical point of view.

The numerical simulation has been done on the {\sf APE}
supercomputer~\cite{ape:machine}.
The model operating at Pisa is the so called ``{\sf tube}'' machine,
a 128 processor parallel computer with a peak performance of 6 GigaFlops.
All the code has been written in the high-level
``{\sf apese}'' language~\cite{ape:language}.

We have used the Conjugate Gradient algorithm
(see~\cite{hest,prossi} for a discussion oriented to the Dirac
lattice operator)
to obtain iteratively the inverse
of the sparse symmetric matrix $A = M^T(\sigma) M(\sigma)$.
We did not use any kind of preconditioning, the improved Symanzik action has
next-to-neighbours interactions and an efficient incomplete
factorization is not trivial.
Moreover, the condition number of $A$ becomes larger when $\Sigma\to 0$,
but at
$\lambda=2.0$ on the $40^2$ lattice, the average $\Sigma$
(which is the fermion mass) and
its fluctuations provide a good rate of convergence and the
inversion problem was not critical.
We tried also to invert directly the matrix $M$ using the Biconjugate
Gradient algorithm~\cite{biconj} but we did not see any advantage.
The stopping condition in the Conjugate Gradient algorithm is
usually chosen to be a bound like
\begin{equation}
||r|| \le \varepsilon_{CG} ||b||
\end{equation}
$r$ is the residual vector: in the following we shall denote with $r_n$ the
recursive residual computed on fly by the algorithm and we shall denote
with $\tilde{r}_n$
the residual defined by
\begin{equation}
\tilde{r}_n = b - A x_n.
\end{equation}
We stress an important fact: let $x^*$ be the exact solution,
the Conjugate Gradient algorithm guarantees that $||x_n - x^*||$
monotonically decreases. On the
other hand $||r_n||$ and $||\tilde{r}_n||$ may show relatively large
fluctuations.
Hence a decreasing trend must be checked
on the succession of residuals to keep under control this effect.
Let us now address the problem of numerical precision.
In absence of rounding errors there would be no difference at all between
the two residuals $r_n$ and $\tilde{r}_n$ as
can be checked inductively using the explicit form of the algorithm.
The effect of
rounding errors associated to a finite precision machine can be easily
studied on simple model
problems (like harmonic problems on regular lattices). The result is that,
given the exact
solution $x^*$, the recursive residual $r_n$ tends to zero, while the true
error has a
positive infimum
\begin{equation}
\frac{||x_n-x^*||}{||x^*||}\to\varepsilon\quad\mbox{as}\quad n\to\infty
\end{equation}
of the order of the machine precision. The small number $\varepsilon$ may
be further reduced
if all the scalar products in the algorithm are implemented as binary sums.
The same saturation phenomena occurs to the
residual $\tilde{r}$, namely
\begin{equation}
||\tilde{r}_n||\to\varepsilon^\prime\ ||b||\quad\mbox{as}\quad n\to\infty .
\end{equation}
Here $\varepsilon^\prime$ is a number which depends on the machine word
length, but also
on the norm of $A$. Typically the saturation of $||x_n-x^*||/||x^*||$
and $||\tilde{r}_n||$
happens roughly at the same time. In a realistic problem $x^*$ is not
known and one can study
the behaviour of $r_n$ and $\tilde{r}_n$.
The departure of $r_n$ from $\tilde{r}_n$ is typically very slow until
the machine
precision is reached and saturation sets up. Our stopping condition has been
\begin{equation}
||r_n|| < 10^{-8}
\end{equation}
and we checked on configurations taken randomly during the runs that the
saturation of
$||\tilde{r}_n|| $ was reached.

We used a $40^2$ lattice with $N=10$ flavours.
In the framework of the $1/N$ expansion one can study the width of the
scaling window,
which is roughly measured by looking at the value of $\Sigma$ in
the crossover region where the weak and strong coupling expressions for
the perturbative mass
are equal. A detailed discussion of the scaling properties of the model
including next-to-leading
$1/N$ corrections and finite size effects can be found in~\cite{curci}.
We have chosen $\lambda = 2.0$ at which the correlation length is about three
lattice spacings and we are in the scaling window.
The volume factor
in favor of the improved Symanzik action is about $15$ and at $L=40$ we
expect from
the finite lattice $1/N$ prediction to have very small
finite size effects.

We determined the critical Wilson mass at which a first order transition
takes place
using mixed-phase runs and measured in
the two definite phases $\langle\Sigma\rangle$ and the composite
$\langle \Sigma^2\rangle_c$ which provides informations on the steepness of
the vacua.
For these quantities the next-to-leading $1/N$ corrections~\cite{curci}
can be computed and
compared with the numerical simulation.

On the $40^2$ lattice the theoretical predictions are $m_c = -0.974$ and
\begin{eqnarray}
\langle \Sigma_-\rangle   &=&  -0.293, \qquad\qquad
\langle \Sigma^2_-\rangle_c =   0.190, \qquad\qquad \\
\langle \Sigma_+\rangle   &=& \phantom{-}  0.311, \qquad\qquad
\langle \Sigma^2_+\rangle_c =   0.146\ . \qquad\qquad
\end{eqnarray}
We started taking $k=1$. After a very rough tuning of $\gamma$ and $\epsilon$
using definite phase runs with the $1/N$ theoretical mass, we
choose $\gamma = 5.0$ and
$\epsilon = 0.06$ for the mixed-phase runs.
Setting half of the lattice to the
positive vacuum and the other half to the negative one, after a short
thermalization, a kink
configuration appears which is very long-lived when the bare mass is at
its critical value.
{}From inspection of the leading effective potential one checks that bare
masses smaller than the
critical one favours the negative vacuum and viceversa.
In Fig.(I)
we show the time evolution of $\langle \Sigma\rangle$ for 16 different
values of
the bare mass starting from $m_c = -0.958$ separated by
$\Delta m_c = 0.002$. The result is
\begin{equation}
m_c = -0.969\pm 0.002,
\end{equation}
In this part of the simulation we did not insist on a heavy tuning of
$\gamma$, $\epsilon$ and $k$
because we were mainly interested in the exact determination of $m_c$.
We then turned to a detailed tuning
of $\gamma$ and $\epsilon$ using definite phase runs at the critical point.
During these runs $k$ was set to the unoptimized value $6$.
In Tab.~(\ref{horowitz_table}) we report measures of autocorrelations
obtained in the two phases of the model with different values
of $\gamma$ and the common choice $\epsilon = 0.09$.
$P_{acc}$ was about $70$--$75\%$ and greater values $\epsilon$ did show a
sharp decrease
in $P_{acc}$.
We did not find any sistematic dependence of $\gamma$ for the two measured
expectation values
of $\langle\Sigma\rangle$ and $\langle\Sigma^2\rangle_c$.

All autocorrelations are computed with the automatic windowing
algorithm~\cite{sokal}
by varying the $c$ constant in the range $4.0$--$8.0$, cross-checking the
result with the
statistical inefficiency~\cite{si} and using consistently at least
$1500\tau$ sweeps.

In the mixed-phase, unoptimized Hybrid Monte Carlo runs gave the same
value for the critical
mass. We did not compare the performances because optimization in
mixed-phase runs was too
time consuming. In the definite phase runs we have taken $\epsilon = 0.06$
and
$N_{md} = 8$ corresponding to a ratio of $\cal C$ factors between the two
algorithms of about $7.7$.
The related rescaled autocorrelations are shown in Tab.~(\ref{hmc_table}).
The resulting expectation values were always consistent with the Kramers
algorithm and the $1/N$ predictions.

Best estimates for the average values extracted only from the runs with the
Kramers algorithm
at different $\gamma$ are
\begin{eqnarray}
\langle \Sigma_-\rangle   &=&  -0.279 \pm 0.002,  \qquad\qquad
\langle \Sigma^2_-\rangle_c = 0.189 \pm 0.001,  \qquad\qquad \\
\langle \Sigma_+\rangle   &=&  \phantom{-}0.316 \pm 0.002,  \qquad\qquad
\langle \Sigma^2_+\rangle_c = 0.145 \pm 0.001,  \qquad\qquad
\end{eqnarray}
and are compatible with both the theoretical predictions and the
Hybrid Monte Carlo results.

{}From the data obtained we see that in the negative phase $\gamma_-=1.0$ is
the optimal value for
both the operators. The same happens in the positive phase but
with a smaller optimal value
$\gamma_+ = 0.5$.
We can see that in the negative phase the performances of the two
algorithms are comparable,
whereas in the positive one the Kramers algorithm is about $1.5$ times slower.

The existence of an optimal range for the parameter $\gamma$ at
fixed $\epsilon$ is easily
understood as follows. In the $\gamma\to\infty$ the Hybrid Monte Carlo
with $N_{md}=1$ is recovered
and it is well known that this is far from being an optimized regime.
In the opposite
limit, $\gamma\to 0$, we obtain the classical equations of motion in the
background $\chi$ field between two refreshments. Of course molecular
dynamics without noise
is highly self-correlated and again a large $\tau$ results. In the
Gross-Neveu model the
$\gamma\to 0$ limit is particularly dangerous. We can study the
$N=\infty$ model taking for
$m_c$ the theoretical prediction including next-to-leading $1/N$
correction on the $40^2$ lattice
at $\lambda=2.0$. One finds that the $\Sigma$ effective potential is so
unbalanced that one of the
two minima disappears. In other words, if the quantum (noise) dynamics is
switched off keeping
all the parameters fixed, the behaviour of the system changes dramatically.
\section{Summary and conclusions}
\label{summary}
This work is a first numerical application of a previous investigation
on the Kramers
equation approach to the lattice simulations of field theoretical models.
We proposed many schemes among which
the simplest (introduced in~\cite{horowitz2}) is studied here.
The situation which we have chosen should be one of the worst possible
cases, namely a
model without bosonic degrees of freedom and with a dynamical symmetry
breaking mechanism
which needs a peculiar determination of the critical point. The main
point of our simulation
is that a not so critical tuning of its parameters gives a performance
comparable to that of
the Hybrid Monte Carlo. This
is quite important with an eye on realistic models, like $QCD$, where a
precise tuning is
too expensive to be done. Moreover, the numerical precision argument
makes us confident
that the present algorithm can be valuable as an independent check for
eventual biases
of the Hybrid Monte Carlo.
Simple scaling arguments (possibly  naive because of the peculiar
negation of momenta)
indicate that the proposed algorithm should improve its
performances with greater volumes or smaller fermion masses.
\acknowledgments
We thank Prof. Paolo Rossi for a careful reading of the manuscript and
many useful discussion.
We thank Prof. Raffaele Tripiccione for a continuous help in developing
the codes for the {\sf APE}
machine.
We also acknowledge the support of many people from the Rome {\sf APE}
group, particularly Simone Cabasino and Gian Marco Todesco.

R. Gupta, G. W. Kilcup and S. R. Sharpe,
Phys. Rev. D {\bf 38}, 1278 (1988).

S. Gupta, A. Irback, F. Karsch and B. Petersson,
Phys. Lett. {\bf 242}B, 437 (1990).
\bibitem{horowitz1}
A. M. Horowitz,
Phys. Lett. {\bf 156} B, 89 (1985).

A. M. Horowitz,
Nucl. Phys. B{\bf 280}, 510 (1987).
\bibitem{horowitz2}
A. M. Horowitz,
Phys. Lett. {\bf 268} B, 247 (1991).
\bibitem{curci}
M. Campostrini, G. Curci and P. Rossi,
Nucl. Phys. B{\bf 314}, 467 (1989).
\bibitem{expansion}
J. F. Schonfeld,
Nucl. Phys. B{\bf 95}, 148 (1975).

R. G. Root,
Phys. Rev. D{\bf 11}, 831 (1975).
%
%
%
%
%
%
\bibitem{rossi}
M. Campostrini and P. Rossi,
Nucl. Phys. B{\bf 329}, 753 (1990).
\bibitem{rosenstein}
B. Rosenstein, B. J. Warr, S. H. Park,
Phys. Rep. {\bf 205}, 59 (1991).
\bibitem{counterterm}
C. Luperini and P. Rossi,
Ann. Phys. {\bf 212}, 371 (1991).
\bibitem{symanzik}
K. Symanzik, in {\sl Mathematical Problems in theoretical Physics}, Lecture
notes in physics {\bf 153}, ed. R. Schrader et al. (Springer, Berlin, 1983).
\bibitem{biconj}
W. H. Press and S. A. Teukolsky,
{\sl Biconjugate Gradient Method for Sparse Linear Systems} in
Comp. in Phys. {\bf 6}, 400 (Jul/Aug 1992).
\bibitem{hest}
M. Hestenes and E. Stiefel,
Nat. Bur. of Standards Journ. Res. {\bf 49}, 409 (1952).
\bibitem{prossi}
P. Rossi, C. T. H. Davies and G. P. Lepage,
Nucl. Phys. B{\bf 297}, 287 (1988).
\bibitem{ape:machine}
A. Bartoloni et al.,
An hardware implementation of the APE100
architecture,
Int. Journ. of Mod. Phys. C, (1993) in press.
\bibitem{ape:language}
A. Bartoloni et al.,
The software of the APE100 processor,
Int. Journ. of Mod. Phys. C, (1993) in press.
\bibitem{sokal}
A. Sokal,
Monte Carlo Methods in Statistical mechanics: Foundations and New Algorithms,
{\sl Cours de Troisi\`eme Cycle de la Physique en Suisse Romande}.

N. Madras and A.D. Sokal,
Jour. of Stat. Phys. {\bf 50}, 109 (1988).
\bibitem{si}
R. Friedberg and J. E. Cameron,
J. Chem. Phys. {\bf 52}, 6049 (1970)
%
%
%
%
\narrowtext
\begin{table}
\caption{{\bf Horowitz} $\epsilon = 0.09$ $k=6$}
\vskip 0.5truecm
\begin{tabular}{ccc}
\label{horowitz_table}
$(\Sigma_0, \gamma)$ & $\tau_\Sigma$ & $\tau_{\Sigma^2}$ \\
\tableline
($-$, 0.2) & 53(7) & 42(6) \\
($-$, 0.5) & 49(5) & 40(4) \\
($-$, 1.0) & 44(4) & 37(3) \\
($-$, 2.0) & 58(7) & 46(5) \\
($-$, 5.0) & 64(8) & 47(5) \\
($+$, 0.2) & 43(5) & 35(4) \\
($+$, 0.5) & 38(3) & 29(3) \\
($+$, 1.0) & 44(5) & 31(3) \\
($+$, 2.0) & 41(4) & 29(2) \\
($+$, 5.0) & 66(9) & 41(5) \\
\end{tabular}
\end{table}
\begin{table}
\caption{{\bf Hybrid} (rescaled by $\cal C$) $\epsilon = 0.06$ $N_{md}=8$}
\vskip 0.5truecm
\begin{tabular}{ccc}
\label{hmc_table}
$\Sigma_0$ & $\tau_\Sigma$ & $\tau_{\Sigma^2}$ \\
$-$ & $44(3)$ & $32(2)$ \\
$+$ & $26(2)$ & $19(1)$ \\
\end{tabular}
\end{table}
\end{document}